\documentclass[%
reprint,
 amsmath,amssymb,
 aps,
prb,
]{revtex4-2}

\usepackage{graphicx}% Include figure files
\usepackage{dcolumn}% Align table columns on decimal point
\usepackage{bm}% bold math
\usepackage{hyperref}
\usepackage{subfigure}
\usepackage{amsfonts}
\usepackage{multirow}% http://ctan.org/pkg/multirow
\usepackage{booktabs}% http://ctan.org/pkg/booktabs

\begin{document}

\title{Diffusion Models for Conditional Generation of Hypothetical \\ New Families of Superconductors}
\date{\today}

\author{Samuel Yuan}
    \email{sdkyuan@gmail.com}
\affiliation{Homestead High School, Cupertino, CA 95014, USA}
 
 \author{S.V. Dordevic}
    \email{dsasa@uakron.edu}
\affiliation{Department of Physics, The University of Akron, Akron, OH 44325, USA}

\begin{abstract}
Effective computational search holds great potential for aiding the discovery of High-Temperature Superconductors (HTSs), especially given the lack of systematic methods for their discovery. Recent progress has been made in this area with machine learning, especially with deep generative models, which have been able to outperform traditional manual searches at predicting new superconductors within existing superconductor families but have yet to be able to generate completely new families of superconductors. We address this limitation by implementing conditioning---a method to control the generation process---for our generative model and develop SuperDiff, a Denoising Diffusion Probabilistic Model (DDPM) with Iterative Latent Variable Refinement (ILVR) conditioning for HTS discovery---the first deep generative model for superconductor discovery with conditioning on reference compounds. With SuperDiff, by being able to control the generation process, we were able to computationally generate completely new families of hypothetical superconductors for the very first time. Given that SuperDiff also has relatively fast training and inference times, it has the potential to be a very powerful tool for accelerating the discovery of new superconductors and enhancing our understanding of them.
\end{abstract}

\maketitle

%%% TODO: INSTEAD OF REF USE et al for all
% TODO: Update Figures correct captions
% Add code/data availability and figure out how to comply with format - see https://arxiv.org/pdf/2302.08882 for format example

\section{Introduction}
\label{introduction}

Superconductors exhibit zero resistivity and perfect diamagnetism. These traits lend them useful for various important technologies, including Maglev trains, MRI magnets, power transmission lines, and quantum computers. However, a major current limitation is that the superconducting transition temperatures ($T_c$) of all known superconductors at ambient pressures are well below room temperature, restricting their broader practical application. Consequently, the search for superconductors with higher $T_c$ is a very active field, as they have significant potential to considerably improve the efficiency of current technologies while also enabling new ones.

Currently, however, superconductivity in high $T_c$ superconductors is not very well understood. As a result, there exists no systematic method for searching for new high $T_c$ superconductors \cite{HIRSCH20151}, and the most common method for searches for new high $T_c$ superconductors is essentially trial-and-error. For instance, the study in Hosono \textit{et al.}~\cite{Hosono2015} surveyed approximately 1000 compounds over four years, of which they found only about $3\%$ to be superconducting. That study is a testament to the extreme inefficiency of finding new high $T_c$ superconductors through pure manual search.

Understanding this, more recently, computational techniques have been applied to assist researchers in the search for new high $T_c$ superconductors. Specifically, a number of works have applied machine learning to this search for superconductors. Although serving as very valuable tools in many respects, most of these attempts \cite{Stanev2018, ROTER20201353689, PhysRevB.103.014509}, have been limited to classification and regression models, which only search through existing databases and are not able to generate any new compounds. Only recently, with deep generative models applied to superconductor discovery, have new hypothetical superconductors not found in most popular compound datasets been generated \cite{Kim_2024, wines2023cdvae, zhongdiffsupercon}. In Kim and Dordevic~\cite{Kim_2024}, a Generative Adversarial Network (GAN) \cite{goodfellow2014} was applied for unconditional high $T_c$ superconductor generation, and in Wines \textit{et al.}~\cite{wines2023cdvae}, a Crystal Diffusion Variational Autoencoder (CDVAE) \cite{xie2021crystal} was also applied for unconditional superconductor generation so that crystal structure could be accounted for; however, that work used a different dataset and focused on the different task of generating stoichiometric Bardeen–Cooper–Schrieffer (BCS) conventional superconductors \cite{PhysRev.106.162} and so did not generate any superconductors with $T_c \gtrsim$~20~K.

New attempts at high $T_c$ superconductor discovery with generative models are not without limitations, however. Most notably, although past models have been able to successfully generate new superconductors within existing superconductor families, they have not been able to generate completely {\it new families} of superconductors, which would be particularly desirable. This is because they are only unconditional models, which learn only the training dataset distribution. As unconditional models, the generation process of these models cannot be controlled. In other words, past models lack conditioning functionality---a method for controlling the generation process, that, in this context, means giving an example superconductor, the reference compound, and having the model generate similar superconductors, ideally by interpolating between the example and what the model has learned from the training dataset. With conditioning, the possibility of generating new families of superconductors can be opened, and researchers can be given control over the generation process. This can be especially useful for researchers looking to find only specific types of superconductors or expand on their own new discoveries. Parallel to our work, Zhong \textit{et al.}~\cite{zhongdiffsupercon} also applied a diffusion model for high $T_c$ superconductor discovery; however, their model was, like previous GANs, greatly limited by its lack of support for conditional generation with reference compounds---which is our main focus. Thus, their diffusion model shared with previous models the major limitation of being unable to generate any {\it new families} of superconductors---essentially, their work was only recreating the performance of the GAN in Kim and Dordevic~\cite{Kim_2024} but with a diffusion model instead and added $T_c$ label control only. Once again, we note that, in this work, we consider ``conditioning'' to mean conditioning the model on reference compounds only, as only this allows for the controlled generation of known and new families of superconductors. Moreover, Kim and Dordevic~\cite{Kim_2024} also struggled at generating unique (distinct from others in the given generated set) pnictides because of the small number of pnictides in SuperCon, the training dataset.

To resolve these limitations, in this work, we implement a Denoising Diffusion Probabilistic Model (DDPM) \cite{ho2020denoising, pmlr-v37-sohl-dickstein15} for superconductor generation as our unconditional model and further implement conditioning with the Iterative Latent Variable Refinement (ILVR) \cite{choi2021ilvr} extension to DDPM, which allows for one-shot generation without additional training. With conditioning, we hope to be able to generate {\it new families} of superconductors for the first time, as identified by the clustering analysis proposed in Roter \textit{et al.}~\cite{ROTER20221354078}, by experimenting with feeding the model different reference superconductors---this would mark a leap in the capabilities of computational searches for superconductors.

Diffusion models are a class of deep generative models that are inspired by nonequilibrium thermodynamics \cite{pmlr-v37-sohl-dickstein15} and have recently shown superior performance and outperformed GANs in image synthesis \cite{dhariwal2021diffusion} and materials discovery \cite{alversonGANandDiff2022}. Diffusion Models are also at the heart of popular new image generation software, such as DALL·E 2 \cite{ramesh2022hierarchical} and Stable Diffusion \cite{rombach2021highresolution}. More recently, these models have also been implemented and shown considerable promise for a variety of scientific applications, such as for drug discovery \cite{corso2023diffdock}.

We coin this first approach to conditionally generating new superconductors with reference compounds as ``SuperDiff''. With SuperDiff, we aim to resolve the issues found in past works as a result of the small pnictide training dataset with the unconditional DDPM and, as our main focus, explore how the conditional DDPM can adapt to new information to generate completely new families of superconductors for the first time.

\section{Methodology}
\label{methodology}

As stated in the introduction, we leverage the capabilities of Denoising Diffusion Probabilistic Models and Iterative Latent Variable Refinement to propose a method for conditionally generating new hypothetical superconductors. Here, we discuss the details of the creation of SuperDiff by discussing the sourcing and processing of superconductor data, providing a brief overview of the details of the underlying DDPM and ILVR methods used, and discussing the techniques we use to evaluate the quality of SuperDiff outputs.

\subsection{Data Processing}
\label{sec:dataprocess}

\begin{figure}
    \centering
    \includegraphics[width = 5.8cm]{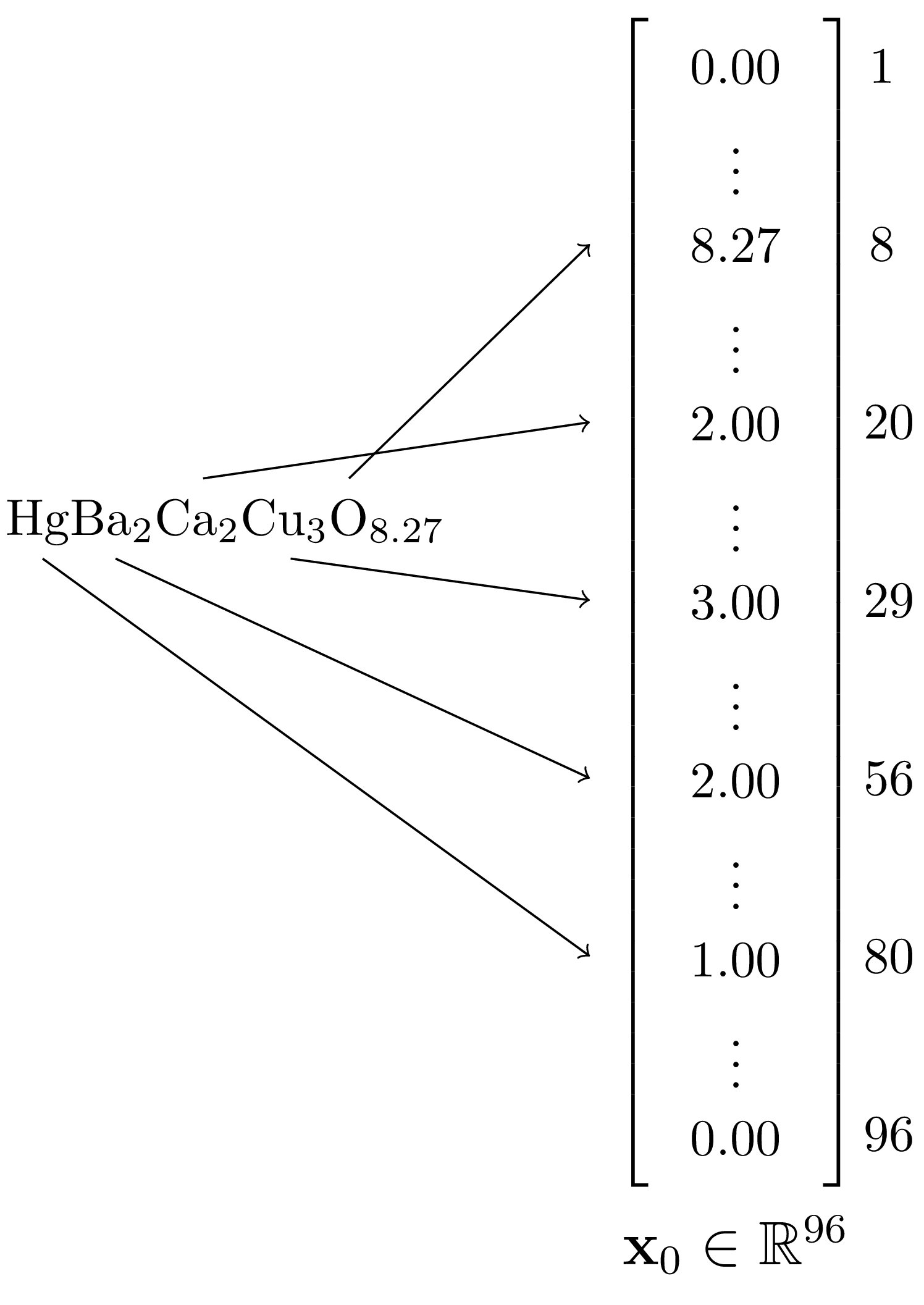}
    \caption{The column vector encoding method used. The figure shows the chemical composition of $\mathrm{HgBa_2Ca_2Cu_3O_{8.27}}$ being encoded as a vector in $\mathbb{R}^{96}$ which is fed to the diffusion model as $\mathbf{x}_0$.}
    \label{fig:compoundencoding}
\end{figure}

All data for the model was sourced from SuperCon\cite{supercondataset}, which is the largest database for superconducting materials. The dataset was processed by the steps in Kim and Dordevic~\cite{Kim_2024} and, like in previous studies \cite{ROTER20201353689,ROTER20221354078,Kim_2024}, only the chemical composition data was used. Every compound from SuperCon was represented as a column vector for input into the model. As shown in Fig. \ref{fig:compoundencoding}, each compound was encoded as a $96 \times 1$ column vector as $96$ is the maximum atomic number present in the dataset.

\subsection{Denoising Diffusion Probabilistic Model}
\label{sec:ddpm}

\begin{figure*}
        \centering
        \includegraphics[width = 14cm]{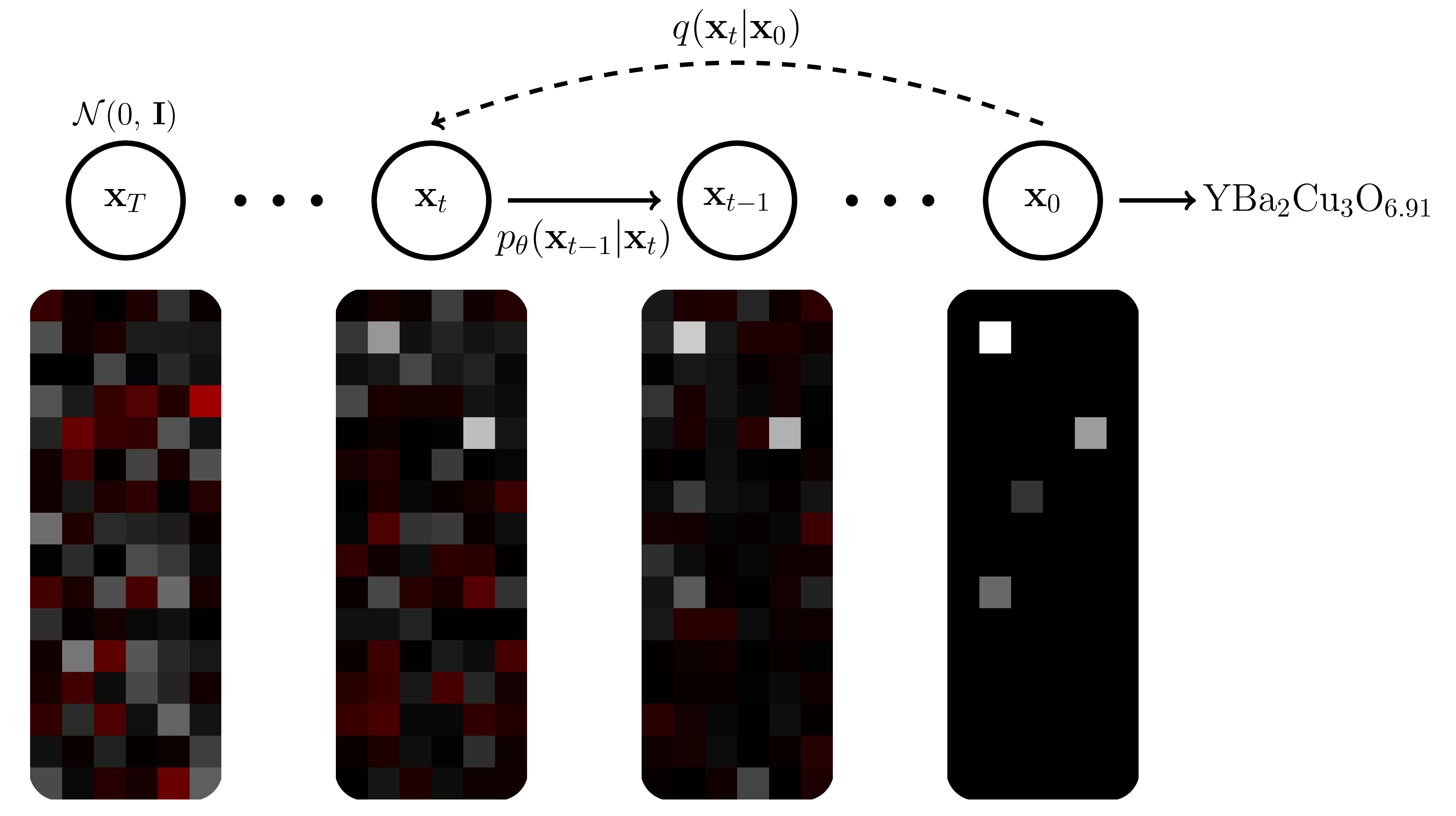}
        \caption{Overview of the unconditional DDPM used. Compounds are encoded as vectors in $\mathbb{R}^{96}$; however, for illustration purposes, the vectors are represented as $16 \times 6$ pixel images in this figure, where each pixel in the image represents an element of the vector, starting from the top-left corner and proceeding horizontally row by row. Whiter pixels represent more positive values (all values are divided by the maximum element of $\mathbf{x}_0$), and redder pixels represent more negative values (black is zero). Starting from noise $\mathbf{x}_T$, the model generates a compound $\mathbf{x}_0$ by denoising $\mathbf{x}_t$ iteratively. Note that $\mathrm{YBa_{2}Cu_{3}O_{6.91}}$ was picked from SuperCon for illustration purposes only, and is not a compound generated by SuperDiff.}
        \label{fig:diffarch}
\end{figure*}

Denoising Diffusion Probabilistic Models (DDPMs) \cite{ho2020denoising, pmlr-v37-sohl-dickstein15} function by learning a Markov chain to progressively transform an isotropic Gaussian into a data distribution. The general structure of the DDPM used is shown in Fig. \ref{fig:diffarch}. The DDPM consists of two parts: a forwards ``diffusion'' process that adds noise to data, and a generative reverse process that learns the reverse of the forwards process---``denoising'' the forwards process. The forward process is a fixed Markov chain that gradually adds Gaussian noise to data. Each step in the forward process is defined as
\begin{equation}
    q(\mathbf{x}_t | \mathbf{x}_{t-1}) := \mathcal{N}(\mathbf{x}_t; \sqrt{1-\beta_t}\mathbf{x}_{t-1}; \beta_t\mathbf{I})\, ,
\end{equation}
where $\beta_1, ..., \beta_T$ is the variance schedule, $\mathbf{I}$ is the identity matrix, and $\mathbf{x}_0$ is dimensionally equivalent to latent variables $\mathbf{x}_1, ..., \mathbf{x}_T$ (all vectors in $\mathbb{R}^{96}$). In this work, we adopt the cosine variance schedule proposed in Nichol and Dhariwal \cite{nichol2021improved}.

A notable property of the forwards process is that given clean data $\mathbf{x}_0$, noised data at any time-step $\mathbf{x}_t$ can be sampled in closed-form:
\begin{equation}
    q(\mathbf{x}_t | \mathbf{x}_0) := \mathcal{N}(\mathbf{x}_t; \sqrt{\overline{\alpha}_t}\mathbf{x}_{0}; (1-\overline{\alpha}_t)\mathbf{I})\, ,\label{eq:2}
\end{equation}
where $\alpha_t := 1-\beta_t$ and $\overline{\alpha}_t = \prod_{s=1}^{t} \alpha_s$. This can be reparametrized \cite{kingma2022autoencoding} as:
\begin{equation}
    \mathbf{x}_t = \sqrt{\overline{\alpha}_t}\mathbf{x}_0 + \sqrt{1 - \overline{\alpha}_t}\bm{\epsilon}\, ,
\end{equation}
where $\bm{\epsilon} \sim \mathcal{N}(0, \mathbf{I})$ and is dimensionally equivalent to $\mathbf{x}_0$.

The reverse process is then defined to be
\begin{equation}
    p_{\theta}(\mathbf{x}_{t-1} | \mathbf{x}_t) :=  \mathcal{N}(\mathbf{x}_{t-1}; \bm{\mu}_{\theta}(\mathbf{x}_t, t); \sigma_{t}^2\mathbf{I})\, .
\end{equation}
In this work, we fix $\sigma_{t}^2 = \beta_t$. Then, as shown in Ho \textit{et al.}~\cite{ho2020denoising}, by rewriting $\bm{\mu}_{\theta}$ as a linear combination of $\mathbf{x}_t$ and $\bm{\epsilon}_\theta$, a neural network that predicts $\bm{\epsilon}$ from $\mathbf{x}_t$ with input and output dimensions equal to that of the noise it predicts, the reverse process may be rewritten as:
\begin{equation}
    \mathbf{x}_{t-1} = \frac{1}{\sqrt{\alpha_t}} \left( \mathbf{x}_t - \frac{1 - \alpha_t}{\sqrt{1 - \overline{\alpha}_t}}\bm{\epsilon}_\theta(\mathbf{x}_t,t)\right) + \sigma_t\mathbf{z}\, ,
\end{equation}
where $\mathbf{z} \sim \mathcal{N}(0, \mathbf{I})$. 

\begin{figure*}
        \centering
        \includegraphics[width = 16cm]{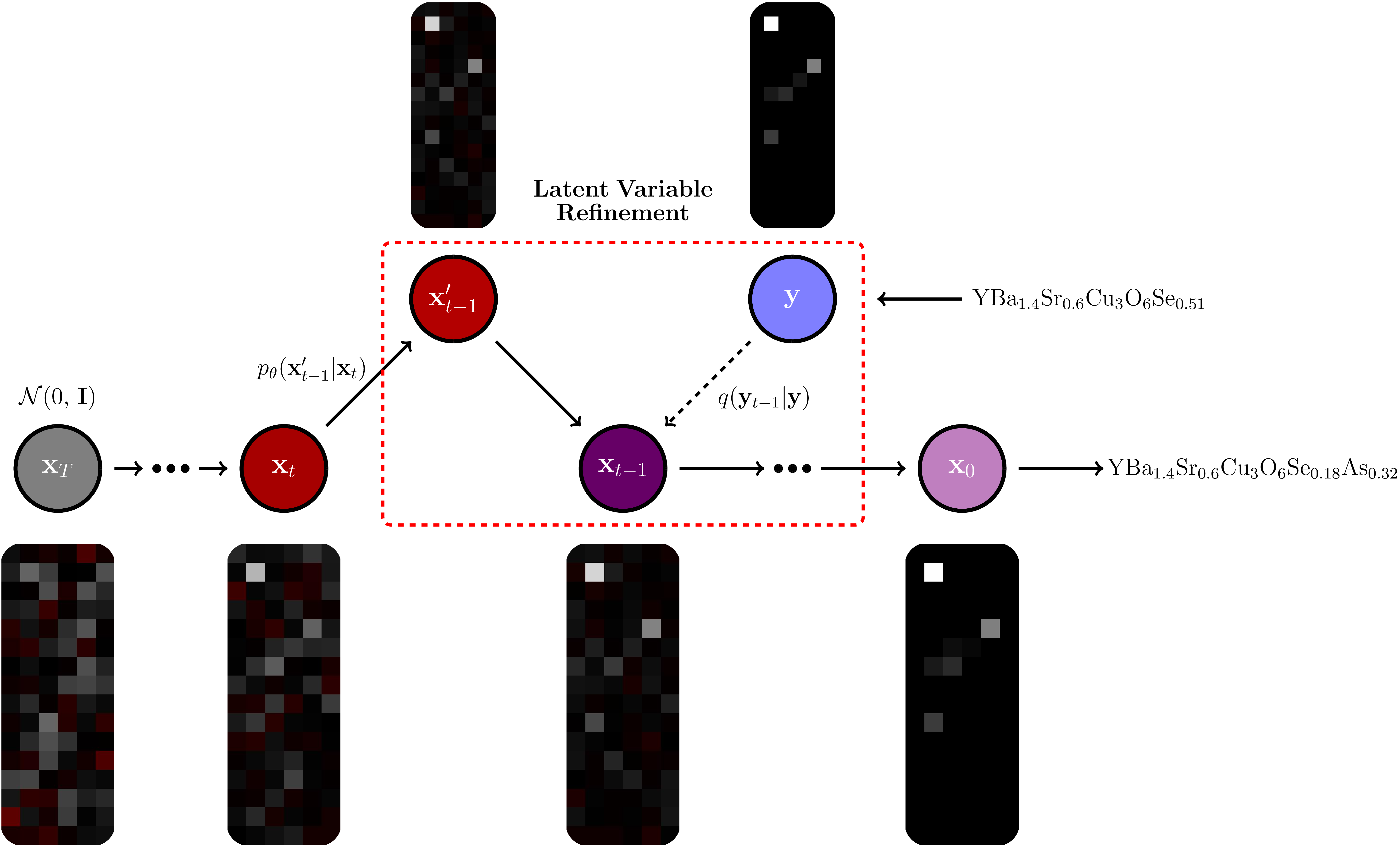}
        \caption{Overview of the Iterative Latent Variable Refinement \cite{choi2021ilvr} method used. The vector image representation is the same as explained in Figure~\ref{fig:diffarch}. $\mathrm{YBa_{1.4}Sr_{0.6}Cu_{3}O_{6}Se_{0.51}}$ \cite{grinenko2023extraordinary} is an example of a reference superconductors and $\mathrm{YBa_{1.4}Sr_{0.6}Cu_{3}O_{6}Se_{0.18}As_{0.32}}$ is an example of a generated output.}
        \label{fig:ilvroverview}
\end{figure*}

To train the DDPM, noise is added to $\mathbf{x}_0$ using the forward process $q(\mathbf{x}_t | \mathbf{x}_{0})$ for a randomly sampled $t \sim \text{Uniform}(\{1, ..., T\})$, which the neural network then learns to remove through the reverse process. 

Four different versions of the DDPM were trained on SuperCon: one for cuprates, one for pnictides, one for others, and one for all classes (``everything''). The training datasets for each version of the DDPM were randomly split into training and validation sets in an approximately $95 \% - 5 \%$ proportion. Training curves for all versions of the DDPM were able to converge and stabilize after around 50 epochs, and each version of the DDPM was trained for between 50 and 100 epochs, depending on the approximate lowest validation loss. For all versions of the DDPM, NAdam \cite{dozat.2016} was chosen as the optimizer, and provided satisfactory results. Moreover, like in Ho \textit{et al.}~\cite{ho2020denoising}, $T$ was set to 1000 and the U-Net \cite{10.1007/978-3-319-24574-4_28} neural network architecture was used for $\bm{\epsilon}_{\theta}$ (for this work, a 1D U-Net was used as opposed to the 2D U-Net used for images).

\subsection{Conditioning}
\label{sec:ilvr}

Iterative Latent Variable Refinement (ILVR) \cite{choi2021ilvr} was used to condition the DDPM. Because ILVR is training-free, the same four trained unconditional DDPMs could be relatively easily modified for conditioning.

ILVR is a slight modification to the reverse diffusion process, and the general structure of ILVR used is shown in Figure.~\ref{fig:ilvroverview}. At each step of the reverse ``denoising'' process, instead of sampling $\mathbf{x}_{t-1}$ directly from $p_{\theta}(\mathbf{x}_{t-1}|\mathbf{x}_t)$ like in unconditional DDPM, $\mathbf{x}_{t-1}$ instead becomes
\begin{equation}
    \mathbf{x}_{t-1} = \phi_{N}(\mathbf{y}_{t-1}) + \mathbf{x}_{t-1}' - \phi_{N}(\mathbf{x}_{t-1}')\, ,
\end{equation}
where $\mathbf{x}_{t-1}' \sim p_{\theta}(\mathbf{x}_{t-1}' | \mathbf{x}_t)$ is the original unconditional proposal, $\mathbf{y}_{t-1} \sim q(\mathbf{y}_{t-1} | \mathbf{y})$ is the condition encoding by the noising process in Equation \eqref{eq:2}, and $\phi_N$ is a linear low-pass filtering operation that maintains the dimensionality of the input.

The goal of ILVR conditioning is to have $\phi_{N}(\mathbf{x}_{0}) = \phi_{N}(\mathbf{y})$, thereby allowing the generated output $\mathbf{x}_0$ to share high-level features with reference $\mathbf{y}$. In this case, the generated superconductor should have similar chemical composition as the reference superconductor.

In Choi \textit{et al.}~\cite{choi2021ilvr}, it was stated that the scale factor $N$ could be changed to control the amount of information brought from the reference to the generated output, where lower $N$ results in greater similarity between generated output and reference and higher $N$ results in only coarse information from the reference being brought by the model to the generated output. In our work, we found that $N > 4$ resulted in large numbers of invalid compounds with negative amounts of elements. As a result, we used $N = 2$ up to $N = 4$, but we still found the conclusions made about changing $N$ in Choi \textit{et al.}~\cite{choi2021ilvr} applicable.

\subsection{Sampling}
\label{sec:sampling}

\begin{table*}
    \centering

    \begin{tabular}{ccccccccccc}
        \hline \hline
        SuperDiff Version & Novel \% & Unique \% & \# Valid & Raw Output \% & True \% Estimate & Mean $T_c$ & SD & Max $T_c$ \\
        \hline
        Everything & 100.00\% & 99.32\% & 79,828 & 67.81\% & 62.22\% & $8.51 \, \mathrm{K}$ & $8.79\,\mathrm{K}$ & $97.0\,\mathrm K$\\
        Cuprates & 100.00\% &  99.92\% & 10,971 & 64.53\% & 58.22\% & $64.68 \, \mathrm{K}$ & $14.61\,\mathrm{K}$ & $110.5 \,\mathrm K$  \\
        Pnictides & 100.00\% & 99.98\% & 2,184 & 95.74\% & 96.39\% & $22.00 \, \mathrm K$ & $2.43 \, \mathrm K$ & $29.5 \, \mathrm K$ \\
        Others & 100.00\% & 99.39\% & 172,739 & 55.82\% & 47.56\% & $6.33 \, \mathrm{K}$ & $2.50 \,\mathrm{K}$ & $30.5\,\mathrm K$  \\
        \hline \hline 
    \end{tabular}

    \caption{Summary of unconditional SuperDiff performance for the four versions we trained from the 500,000 compounds we sampled from each version. Shown are the percentage of generated compounds that were novel (not in the training set) and unique (distinct from others in the given generated set) before \texttt{SMACT} \cite{DAVIES2016617} filters, the number of generated compounds that were valid (passed \texttt{SMACT} filters), the percentages of generated compounds determined to be superconducting by the classification model along with the estimated true percentages according to Eq.~\ref{eq:estimate}, and summary $T_c$ statistics from the predictions by the regression model. We note that although the novelty percentage is 100.00\%, this is due to rounding, and the model does, on extremely rare occasions, exactly reconstruct superconductors from the training dataset.}
    \label{tab:unconditionalverification}
\end{table*}

As mentioned previously, we trained four versions of the unconditional model, each of which was then copied and modified with ILVR conditioning to also create four versions of the conditional model. We thus have four versions of the unconditional DDPM (without ILVR), which we call ``unconditional SuperDiff'', and four versions of the conditional DDPM (with ILVR), which we call ``conditional SuperDiff''. On a single consumer Nvidia RTX 3060 Ti GPU, each version of SuperDiff was trained in under 2 hours, and we sampled 500,000 compounds from each of the four unconditional SuperDiff versions, which took less than 10 hours for each version. These relatively fast training and inference times make SuperDiff a system that can be trained and used using resources at most universities and even consumers. For conditional SuperDiff, we sampled varying amounts of compounds for different reference superconductors, and we discuss those results later.

All sampled compounds were initially screened through various quality checks to ensure that all generated compounds were reasonably realistic. First, we obviously eliminated all generated compounds with negative amounts of elements. Note that we round all amounts of elements to two decimal places beforehand. Next, we eliminate compounds with either too few (only 1) or too many elements---for Cuprates, we limit outputs to compounds with a maximum of 7 elements, and for Pnictides and Others, we limit outputs to compounds with a maximum of 5 elements. After these basic checks, we removed duplicates and further evaluated compound validity with the charge neutrality and electronegativity checks from the \texttt{SMACT} package \cite{DAVIES2016617}. Finally, we ran formation energy prediction with ElemNet \cite{Jha2018, Jha2019}. We will discuss the performance of model generations against these checks later.

\subsection{Clustering}
\label{sec:cluster}

To identify if SuperDiff could generate new superconductor families, clustering analysis was performed. Clustering, which is an unsupervised machine-learning method used to find hidden patterns within data, was applied to the SuperCon database in Roter \textit{et al.}~\cite{ROTER20221354078}, which established that these methods, when applied to superconductors, could exceed human performance in identifying different ``families'' of superconductors, which are represented as clusters. In this work, we use the clustering method for superconductors from Roter \textit{et al.}~\cite{ROTER20221354078} to evaluate generated outputs for new families. In Roter \textit{et al.}~\cite{ROTER20221354078}, it was also found that, for superconductors, to visualize clustering results, the t-SNE method worked best. t-SNE is a non-linear dimensionality reduction technique that allows higher dimensional data ($96$-dimensional superconductor data points in this case) to be represented in 2D or 3D \cite{vanDerMaaten2008} (which do not have any physical meaning).

As discussed in the introduction, a major objective of this work was to generate new families of superconductors, as identified by the clustering model---that is, to generate new clusters of superconductors. This was something not accomplished by previous works, including the GAN in Kim and Dordevic~\cite{Kim_2024} and the diffusion model in Zhong \textit{et al.}~\cite{zhongdiffsupercon}. In order to achieve this goal, we experimented with the conditional model's ability to interpolate between the reference compound and the training dataset. This idea of experimenting with a conditional DDPM's ability to interpolate between the reference set and training set was proposed in Giannone \textit{et al.}~\cite{giannone2022fewshot} to attempt to achieve few-shot generation on image classes never seen during training. We attempt to do this with superconductors in this work. For instance, we experiment with conditioning the cuprate version of conditional SuperDiff on new, different reference cuprates not in the families of cuprate superconductors in the training dataset. We examine the model's ability to generate new clusters or families of superconductors using information from the reference compound with this technique, and we report our clustering results below.

\section{Results}
\label{results}

In this section we report the performance of SuperDiff on various checks and discuss some interesting new findings. We first evaluate the performance of unconditional SuperDiff with the 500,000 compounds we generated for each of the four classes by performing various computational tests, which included some general compound checks as well as checks for superconductivity. We use the same computational tests for unconditional SuperDiff as used for the GAN in Kim and Dordevic~\cite{Kim_2024} and are thus able to directly compare unconditional performance. Afterward, as our most notable results, we evaluate the performance of both the unconditional and conditional versions of SuperDiff on clustering and manually identify and present some promising new families of superconductors generated by the conditional SuperDiff.

\subsection{Duplicates and Validity}
\label{sec:resultsDNV}

\begin{figure*}
    \centering
    \includegraphics[width=15cm]{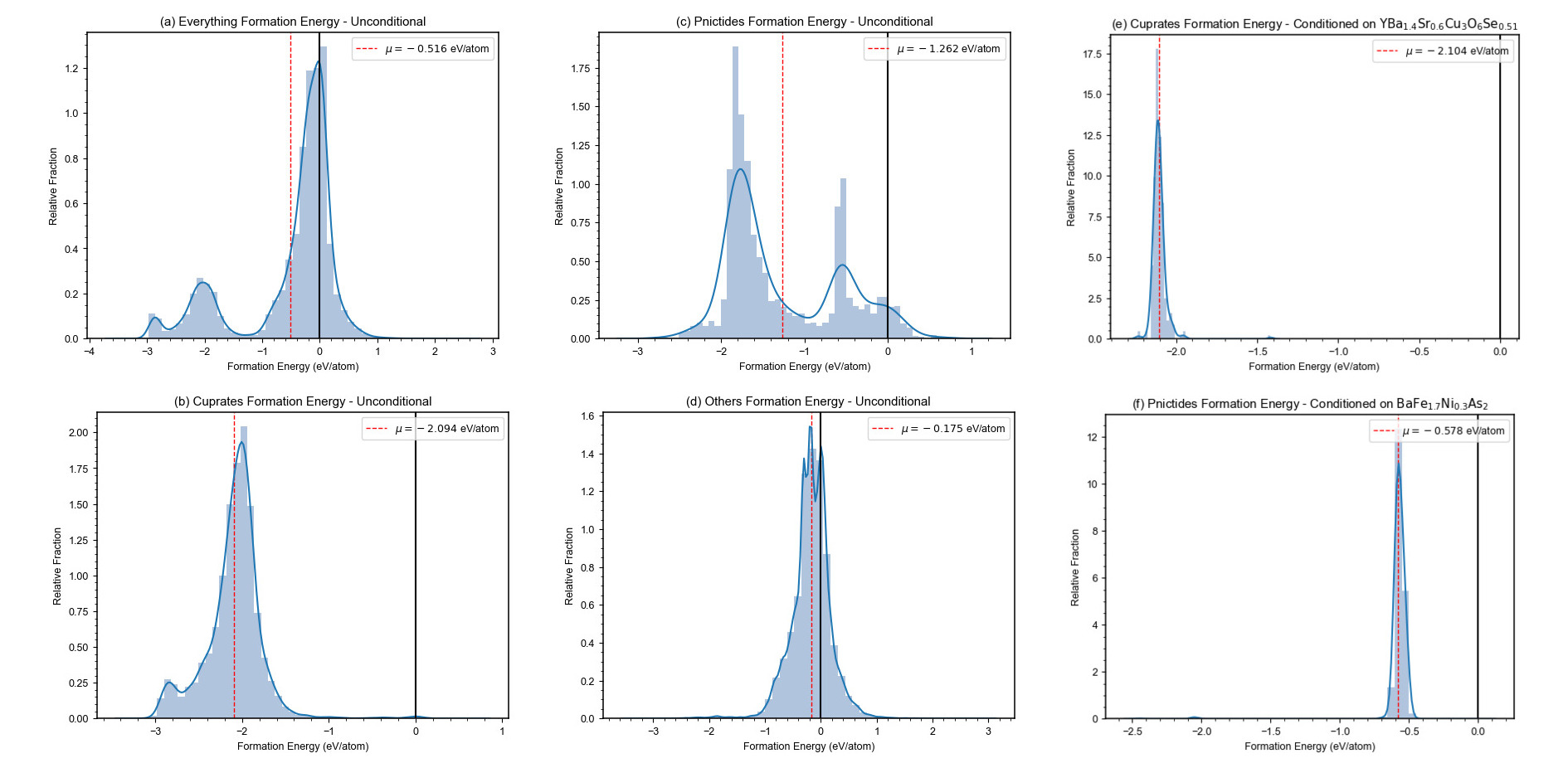}
    \caption{Distribution of ElemNet \cite{Jha2018, Jha2019} predicted formation energies of the generated compounds from the four versions of unconditional SuperDiff---(\textbf{a}) Everything, (\textbf{b}) Cuprates, (\textbf{c}) Pnictides, and (\textbf{d}) Others---as well as (\textbf{e}) Cuprates version of conditional SuperDiff conditioned on $\mathrm{YBa_{1.4}Sr_{0.6}Cu_{3}O_{6}Se_{0.51}}$ \cite{grinenko2023extraordinary} and (\textbf{f}) Pnictides version of conditional SuperDiff conditioned on $\mathrm{BaFe_{1.7}Ni_{0.3}As_{2}}$ \cite{Wang2013}. Also shown are the average formation energy for each distribution.}
    \label{fig:form_energy}
\end{figure*}

For the 500,000 compounds generated by each version of unconditional SuperDiff, we first screened for duplicates between the generated set and the training set (the portion of the SuperCon database of the same class) and for duplicates within the generated set itself. After this, we ran the charge neutrality and electronegativity checks on the generated compounds with the \texttt{SMACT} package \cite{DAVIES2016617}. We present the results of these general tests in Table~\ref{tab:unconditionalverification}, and then we remove all duplicates from the generated sets.

We notice that the novelty \% and uniqueness \% of generated results are all very high, which means that unconditional SuperDiff is able to successfully generate both diverse and novel compounds. Unconditional SuperDiff, here, outperforms the GAN in Kim and Dordevic~\cite{Kim_2024} in all metrics regarding generation novelty and uniqueness, and, similar to as proposed in their work, we also speculate that the high novelty percentage is due to the non-stoichiometric nature of the compounds we generate, which opens up a large composition space for the model. Notably, unconditional SuperDiff maintains a very high uniqueness \% for pnictides despite the small training set, something not accomplished by the Wasserstein GAN in Kim and Dordevic~\cite{Kim_2024}. This corroborates the observation of the superior ability of DDPMs to generate diverse results when compared to a GAN in other disciplines \cite{dhariwal2021diffusion}. Lastly, although the \texttt{SMACT} check \cite{DAVIES2016617} results varied greatly between classes and the proportion of valid compounds for some classes was fairly low, the fast inference time justifies that SuperDiff is still able to generate valid compounds reasonably well for all classes.

Overall, these results indicate that all versions of unconditional SuperDiff are able to generate both novel and unique compounds---overcoming the past issues faced by Kim and Dordevic~\cite{Kim_2024}---as well as valid compounds. As conditional SuperDiff maintains much of the same components as the unconditional model, it was unsurprising that---in most cases---conditional SuperDiff was also able to generate novel, unique, and valid compounds; however, for conditional SuperDiff, these qualities were very much dependent on the reference compound---we still run these checks on all compounds generated by conditional SuperDiff and filter out invalid compounds.

\subsection{Formation Energy}
\label{sec:formationenergy}

We further validated the performance of SuperDiff on generating synthesizable compounds by predicting the formation energies of the generated compounds with ElemNet \cite{Jha2018, Jha2019}, which is a deep neural network model for predicting material properties from only elemental compositions. We chose ElemNet for our formation energy prediction because of its ability to use only chemical composition, as we do not consider crystal structure in our generation process. Because ElemNet does not take in compounds as column vectors in $\mathbb{R}^{96}$, as SuperDiff does, but instead takes them in as column vectors in $\mathbb{R}^{86}$ with certain elements removed, we ran the ElemNet formation energy prediction on only the compounds generated by SuperDiff that ElemNet would directly support---this did constitute the great majority of generated compounds. We display the distributions for the predicted formation energies of generated compounds in Fig.~\ref{fig:form_energy}.

As shown in the figure, unconditional SuperDiff generated a majority of compounds with negative formation energy for all classes of superconductors, with the mean formation energy for all classes predicted to be negative as well. In Jha \textit{et al.}~\cite{Jha2018}, it was stated that negative formation energy values for compounds are a good indicator of their stability and synthesizability; therefore, although these predictions are not definitive proof---experimentation validation would be necessary---these predictions provide an indication that most of the compounds generated by unconditional SuperDiff are plausibly stable and synthesizable.

\begin{figure*}
    \centering
    \includegraphics[width=\textwidth]{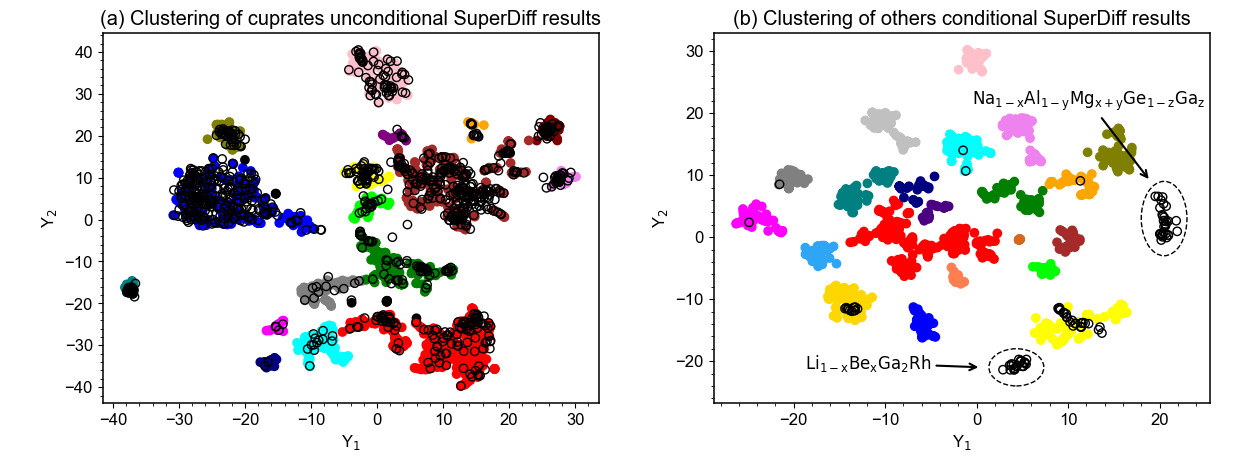}
    \caption{Clustering of the (\textbf{a}) valid generated compounds from the Cuprates version of unconditional SuperDiff and (\textbf{b}) valid generated compounds from the Others version of conditional SuperDiff conditioned on various different compounds. Colored full circles represent data points from SuperCon (cuprates only for (\textbf{a}) and others only for (\textbf{b})), with each color representing a different cluster, or family, of superconductor as identified by the model from Roter \textit{et al.}~\cite{ROTER20221354078}; black open circles are compounds generated by SuperDiff. We notice that unconditional SuperDiff did not generate any new families of superconductors, as all generated compounds fall within the existing clusters of superconductors from SuperCon. However, for conditional SuperDiff, although some generated superconductors fall within the existing SuperCon clusters, we were able to identify two new clusters consisting of only generated superconductors (marked with arrows). These two new clusters correspond to two new families of superconductors generated by SuperDiff: $\mathrm{Li_{1-x}Be_{x}Ga_{2}Rh}$ and $\mathrm{Na_{1-x}Al_{1-y}Mg_{x+y}Ge_{1-z}Ga_{z}}$.}
    \label{fig:combinednewclustering}
\end{figure*}

For conditional SuperDiff, the distribution of formation energies for generated compounds is heavily dependent on the reference compound. However, given a reasonable reference compound---that is, a valid reference compound that belongs to the class of superconductor that the version of SuperDiff was trained on---we demonstrate that conditional SuperDiff is able to generate compounds predicted to be stable by ElemNet. Specifically, as shown in Fig.~\ref{fig:form_energy}, for the cuprates version of conditional SuperDiff conditioned on $\mathrm{YBa_{1.4}Sr_{0.6}Cu_{3}O_{6}Se_{0.51}}$ \cite{grinenko2023extraordinary} and the pnictides version of conditional SuperDiff conditioned on $\mathrm{BaFe_{1.7}Ni_{0.3}As_{2}}$ \cite{Wang2013}---some of the compounds we conditioned conditional SuperDiff on to find new families of superconductors later---the predicted distribution of formation energies for generated compounds show all generated compounds to have negative formation energy. These results indicate that, given reasonable reference compounds, conditional SuperDiff can generate plausibly stable and synthesizable compounds, which is not surprising given the fundamental architecture similarities between conditional and unconditional SuperDiff.

\subsection{Superconductivity}
\label{sec:superconductivity}

\begin{table*}
     \centering
     \begin{ruledtabular}
     \begin{tabular}{lclcl}
     Reference Compound & SuperDiff Version & Output Examples & Predicted $T_c$ & General Formula \\ \hline
         \multirow{3}{*}{$\mathrm{YBa_{1.4}Sr_{0.6}Cu_{3}O_{6}Se_{0.51}}$ \cite{grinenko2023extraordinary}} & \multirow{3}{*}{Cuprates} &  $\mathrm{YBa_{1.4}Sr_{0.6}Cu_{3}O_{6}Se_{0.35}As_{0.12}}$ & $  55 \, \mathrm K$ & 
         \multirow{3}{*}{$\mathrm{YBa_{1.4}Sr_{0.6}Cu_{3}O_{6}Se_{x}As_{y}}$} \\
         & & $\mathrm{YBa_{1.4}Sr_{0.6}Cu_{3}O_{6}Se_{0.28}As_{0.21}}$ & $  41 \, \mathrm K$ & \\
         & & $\mathrm{YBa_{1.4}Sr_{0.6}Cu_{3}O_{6}Se_{0.18}As_{0.32}}$ & $  46 \, \mathrm K$ & \\
         % ... additional rows ...
         % --------------------------
         \hline
         \multirow{3}{*}{$\mathrm{YBa_{1.4}Sr_{0.6}Cu_{3}O_{6}Se_{0.51}}$ \cite{grinenko2023extraordinary}} & \multirow{3}{*}{Cuprates} &  $\mathrm{YBa_{1.4}Sr_{0.6}Cu_{3}O_{6}Se_{0.18}Br_{0.19}}$ & $  54 \, \mathrm K$ & 
         \multirow{3}{*}{$\mathrm{YBa_{1.4}Sr_{0.6}Cu_{3}O_{6}Se_{x}Br_{y}}$} \\
         & & $\mathrm{YBa_{1.4}Sr_{0.6}Cu_{3}O_{6}Se_{0.11}Br_{0.25}}$ & $  33 \, \mathrm K$ & \\
         & & $\mathrm{YBa_{1.4}Sr_{0.6}Cu_{3}O_{6}Se_{0.17}Br_{0.25}}$ & $  41 \, \mathrm K$ & \\
         % ... additional rows ...
         % --------------------------
         \hline      
         \multirow{3}{*}{$\mathrm{SrCu_{2}O_{3}}$ \cite{Ohsugi1999}} & \multirow{3}{*}{Cuprates} &  $\mathrm{SrCu_{1.77}Ni_{0.08}Zn_{0.15}O_{3}}$ & $  31 \, \mathrm K$ & 
         \multirow{3}{*}{$\mathrm{SrCu_{2-x-y}Zn_{x}Ni_{y}O_{3}}$} \\
         & & $\mathrm{SrCu_{1.58}Ni_{0.11}Zn_{0.31}O_{3}}$ & $  10 \, \mathrm K$ & \\
         & & $\mathrm{SrCu_{1.85}Ni_{0.08}Zn_{0.07}O_{3}}$ & $  28 \, \mathrm K$ & \\
         % ... additional rows ...
         % --------------------------
         \hline     
         \multirow{3}{*}{$\mathrm{Ba_2CuO_{3.25}}$ \cite{FUMAGALLI20211353810}} & \multirow{3}{*}{Cuprates} &  $\mathrm{Ba_{1.88}Cs_{0.12}CuO_{3.28}}$ & $  30 \, \mathrm K$ & 
         \multirow{3}{*}{$\mathrm{Ba_{2-x}Cs_{x}CuO_{3.3}}$} \\
         & & $\mathrm{Ba_{1.91}Cs_{0.09}CuO_{3.28}}$ & $  28 \, \mathrm K$ & \\
         & & $\mathrm{Ba_{1.77}Cs_{0.23}CuO_{3.3}}$ & $  13 \, \mathrm K$ & \\
         % ... additional rows ...
         % --------------------------
         \hline      
         \multirow{3}{*}{$\mathrm{LiCu_2O_2}$ \cite{PhysRevB.97.054428}} & \multirow{3}{*}{Cuprates} &  $\mathrm{Li_{0.67}Be_{0.34}Cu_{2}O_{2}}$ & $  33 \, \mathrm K$ & 
         \multirow{3}{*}{$\mathrm{Li_{1-x}Be_{x}Cu_{2}O_{2}}$} \\
         & & $\mathrm{Li_{0.89}Be_{0.11}Cu_{2}O_{2}}$ & $  22 \, \mathrm K$ & \\
         & & $\mathrm{Li_{0.72}Be_{0.28}Cu_{2}O_{2}}$ & $  34 \, \mathrm K$ & \\
         % ... additional rows ...
         % --------------------------
         \hline      
         \multirow{3}{*}{$\mathrm{LiGa_2Rh}$ \cite{MONDAL20221354142}} & \multirow{3}{*}{Others} &  $\mathrm{Li_{0.67}Be_{0.34}Ga_{2}Rh}$ & $  9 \, \mathrm K$ & 
         \multirow{3}{*}{$\mathrm{Li_{1-x}Be_{x}Ga_{2}Rh}$} \\
         & & $\mathrm{Li_{0.87}Be_{0.13}Ga_{2}Rh}$ & $  33 \, \mathrm K$ & \\
         & & $\mathrm{Li_{0.71}Be_{0.29}Ga_{2}Rh}$ & $  27 \, \mathrm K$ & \\
         % ... additional rows ...
         % --------------------------
         \hline      
         \multirow{3}{*}{$\mathrm{NaAlGe}$ \cite{PhysRevMaterials.7.104801}} & \multirow{3}{*}{Others} &  $\mathrm{Na_{0.8}Al_{0.92}Mg_{0.28}Ge_{0.84}Ga_{0.16}}$ & $  10 \, \mathrm K$ & 
         \multirow{3}{*}{$\mathrm{Na_{1-x}Al_{1-y}Mg_{x+y}Ge_{1-z}Ga_{z}}$} \\
         & & $\mathrm{Na_{0.36}Al_{0.63}Mg_{0.99}Ge_{0.88}Ga_{0.12}}$ & $  12 \, \mathrm K$ & \\
         & & $\mathrm{Na_{0.79}Al_{0.75}Mg_{0.46}Ge_{0.68}Ga_{0.32}}$ & $   8 \, \mathrm K$ & \\
         % ... additional rows ...
         % --------------------------
         \hline     
         \multirow{3}{*}{$\mathrm{BaFe_{1.7}Ni_{0.3}As_{2}}$ \cite{Wang2013}} & \multirow{3}{*}{Pnictides} &  $\mathrm{BaFe_{1.72}Co_{0.13}Ni_{0.15}As_{2}}$ & $  24 \, \mathrm K$ & 
         \multirow{3}{*}{$\mathrm{BaFe_{2-x-y}Co_{x}Ni_{y}As_{2}}$} \\
         & & $\mathrm{BaFe_{1.74}Co_{0.08}Ni_{0.08}As_{2}}$ & $  16 \, \mathrm K$ & \\
         & & $\mathrm{BaFe_{1.7}Co_{0.12}Ni_{0.11}As_{2}}$ & $  30 \, \mathrm K$ & \\
         % ... additional rows ...
         % --------------------------
         \hline     
         \multirow{3}{*}{$\mathrm{BaFe_{1.7}Ni_{0.3}As_{2}}$ \cite{Wang2013}} & \multirow{3}{*}{Pnictides} &  $\mathrm{BaFe_{1.84}Co_{0.16}As_{1.8}Ge_{0.2}}$ & $  17 \, \mathrm K$ & 
         \multirow{3}{*}{$\mathrm{BaFe_{2-x}Co_{x}As_{2-y}Ge_{y}}$} \\
         & & $\mathrm{BaFe_{1.77}Co_{0.23}As_{1.81}Ge_{0.19}}$ & $  24 \, \mathrm K$ & \\
         & & $\mathrm{BaFe_{1.82}Co_{0.18}As_{1.79}Ge_{0.21}}$ & $   27 \, \mathrm K$ & \\
         % ... additional rows ...
         % --------------------------

    \end{tabular}
    \end{ruledtabular}
    \caption{Promising new families of superconductors generated by conditional SuperDiff. Shown are the reference compound used to condition the SuperDiff, the version of conditional SuperDiff used, a few output examples from the family and their predicted critical temperatures \cite{ROTER20201353689}, and the general formula for the new family.}
    \label{tab:predictions}
\end{table*}

After those general checks, we performed some computational checks for superconductivity in order to verify that unconditional SuperDiff is indeed able to generate probable superconductors. We ran the compounds generated by unconditional SuperDiff through the $K$-Nearest Neighbors (KNN) classification model and regression model from Roter and Dordevic~\cite{ROTER20201353689} for predicting superconductivity and critical temperature, respectively, based on elemental composition.

For the predicted proportion of generated compounds that were superconducting, we accounted for the inherent probabilistic error of the classification model by using Bayesian statistics to estimate the true proportion of superconducting generated compounds given the classification model's predicted proportion $p_{sc}$ and the true positive $\textit{\textsf{tp}}$ and false positive rates $\textit{\textsf{fp}}$ of the classification model. The true proportion of generated compounds that are superconductors $\rho_{sc}$ may be estimated as \cite{Kim_2024}
\begin{equation}\label{eq:estimate}
    \rho_{sc} \approx \frac{p_{sc} - \textit{\textsf{fp}}}{\textit{\textsf{tp}} - \textit{\textsf{fp}}}\, ,
\end{equation}
where $\textit{\textsf{tp}} = 98.69\%$ and $\textit{\textsf{fp}} = 16.94\%$ are reported by Roter and Dordevic~\cite{ROTER20201353689}.

For the generated compounds that were predicted to be superconducting, we used the regression model in Roter and Dordevic~\cite{ROTER20201353689} to predict their critical temperatures. Like all other tests done so far, this computational prediction is only an approximation. We tabulated the results of the classification and regression predictions on the compounds generated by unconditional SuperDiff in Table~\ref{tab:unconditionalverification}. We will discuss the predicted superconductivity of compounds generated by conditional SuperDiff later.

As seen in the table, all versions of unconditional SuperDiff were able to generate predicted superconductors at a rate comparable to the GAN in Kim and Dordevic~\cite{Kim_2024} and much higher than the 3\% achieved by manual search in Hosono \textit{et al.}~\cite{Hosono2015}---notably, unconditional SuperDiff seems to perform much better on pnictides despite the small training set. This is further indication of the effectiveness of computational search for superconductors when compared to manual searches. Moreover, unconditional SuperDiff seems to capture the critical temperature distribution of the SuperCon training dataset much better than the GAN in Kim and Dordevic~\cite{Kim_2024}.

Although actual synthesis and testing in a lab are required to confirm superconductivity, these checks, combined with the clustering analysis results that we will discuss later, provide a general indication that unconditional SuperDiff is able to generate highly plausible superconductors.

\subsection{Clustering Results}
\label{sec:clusteringresults}

We ran the clustering analysis described previously on both unconditional and conditional SuperDiff. We display the clustering results for the cuprates version of unconditional SuperDiff in Fig.~\ref{fig:combinednewclustering}. Superconductors from the SuperCon database are shown with full circles of different colors, whereas our predictions are shown with open black circles. Although unconditional SuperDiff generated compounds in all known clusters or families of superconductors, no new families of superconductors were generated by unconditional SuperDiff---this was true for the other versions of unconditional SuperDiff as well. This was the expected result for unconditional SuperDiff as the underlying DDPM's goal is to just find a mapping from Gaussian noise to the training data distribution, not some other new distribution. However, superconductor discovery has a particular interest in the generation of new families of superconductors, so a method to control the generation process to change the generated data distribution is desirable. With conditional SuperDiff, we are able to control the generation process to computationally generate new families of superconductors for the first time.

In Fig.~\ref{fig:combinednewclustering}, we also display a sample clustering result from the ``others'' version of conditional SuperDiff conditioned on various compounds. As seen in the plot, we identified two new clusters: $\mathrm{Li_{1-x}Be_{x}Ga_{2}Rh}$, which was generated by conditioning SuperDiff on $\mathrm{LiGa_2Rh}$ \cite{MONDAL20221354142}, and $\mathrm{Na_{1-x}Al_{1-y}Mg_{x+y}Ge_{1-z}Ga_{z}}$, which was generated by conditioning SuperDiff on $\mathrm{NaAlGe}$ \cite{PhysRevMaterials.7.104801}. Those and other predicted families will be discussed in more detail below.

These clustering results show that, with this ability to control generation, and by conditioning SuperDiff on compounds not in the SuperCon training set, SuperDiff is able to use information from various reference compounds to generate completely new families of superconductors. As expected, due to the nature of the conditioning method, we note that for these generated new families, the reference compound does belong to the new cluster generated based on it; however, one of the main contributions presented in this work is that we are able to extrapolate a new family of superconductors from an otherwise single reference compound. We performed this clustering analysis on all versions of conditional SuperDiff conditioned on a variety of different reference compounds, and we discuss the promising new families of superconductors generated by conditional SuperDiff in more detail and verify their superconductivity below.

\subsection{Promising Generated New Families}
\label{sec:newfamilies}

\begin{figure}
% \vspace{-2cm}
    \centering
    \hspace*{-0.5cm}\includegraphics[width=8cm]{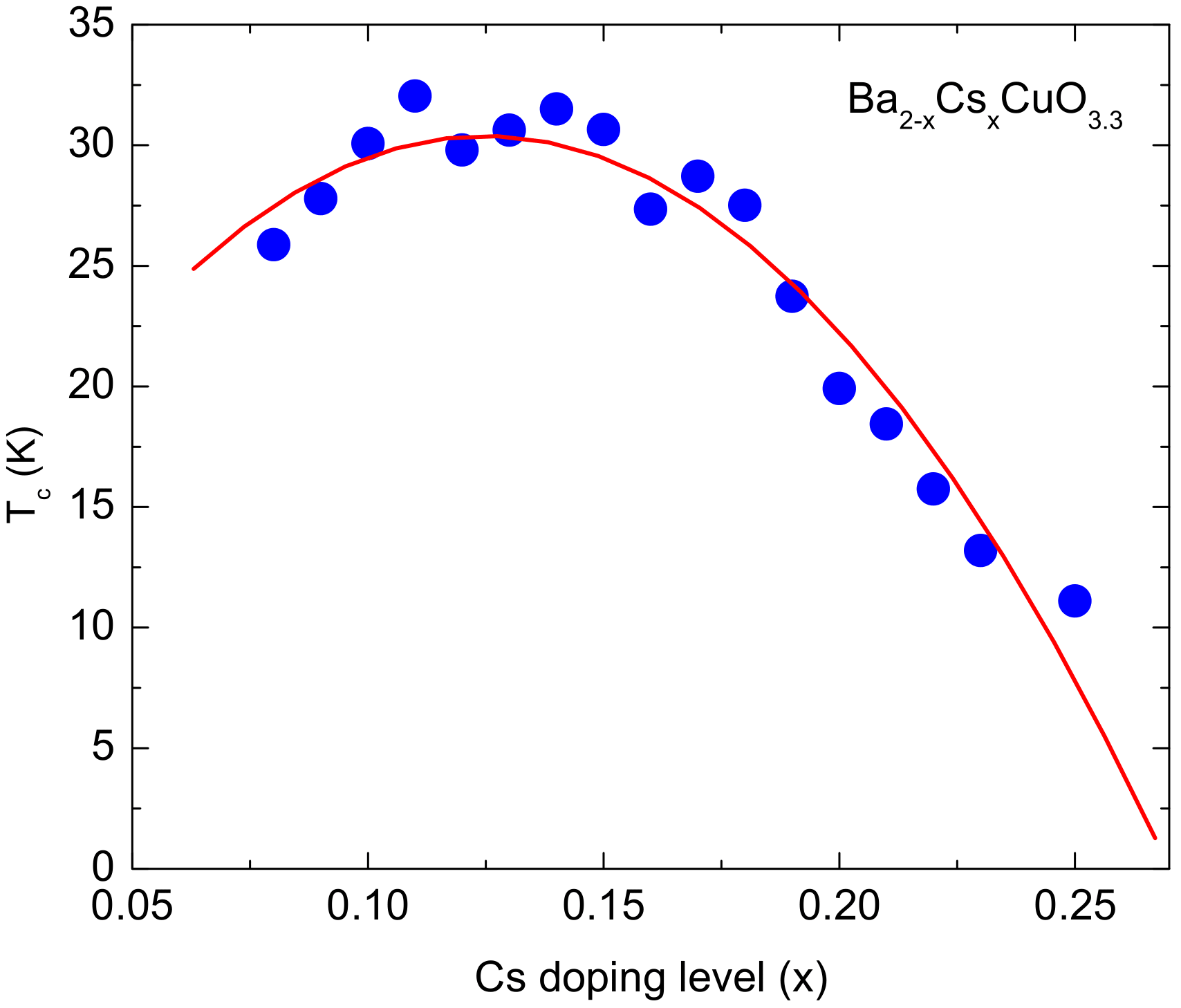}
% \vspace{-2cm}
    \caption{Plot of $T_c$ predicted by the regression model in Roter and Dordevic~\cite{ROTER20201353689} versus Cesium content ($x$) for $\mathrm{Ba_{2-x}Cs_{x}CuO_{3.3}}$ family generated by conditional SuperDiff (see Table~\ref{tab:predictions}). We notice a characteristic parabolic dependence of $T_c$ versus doping, observed previously in other cuprate families \cite{TALLON200153}.}
    \label{fig:doping}
\end{figure}

After running clustering analysis for the different versions of conditional SuperDiff conditioned on a variety of reference compounds, we manually identified the most promising new families of superconductors generated by conditional SuperDiff. Beyond the novelty, uniqueness, and \texttt{SMACT} checks, we further checked for the novelty of these newly generated families by searching on the internet and through other databases---these newly generated families could not be found anywhere else. We tabulated these most promising new families of superconductors generated by conditional SuperDiff in Table~\ref{tab:predictions}. There, we identified the reference compound used as well as a few output examples and their respective predicted $T_c$ using the regression model in Roter and Dordevic~\cite{ROTER20201353689}, and determined the general formula for the new family. We notice that most compounds generated with conditional SuperDiff are predicted to be superconducting, with predicted $T_c$ being reasonable for each class. Additionally, a particularly interesting result to note was that our model seemed to generate some new families of superconductors with double or, in one case, even triple doping. This is an interesting new avenue for superconductor discovery that has not been extensively studied, which our model suggests should be explored in more detail by material scientists.

We further demonstrate that conditional SuperDiff is able to generate realistic new families of superconductors by plotting the predicted $T_c$ using the regression model in Roter and Dordevic~\cite{ROTER20201353689} against the Cesium doping content for the newly generated $\mathrm{Ba_{2-x}Cs_{x}CuO_{3.3}}$ family in Fig.~\ref{fig:doping}. We notice that the generated $\mathrm{Ba_{2-x}Cs_{x}CuO_{3.3}}$ family is predicted to exhibit the expected parabolic $T_c$ doping dependence relationship for this type of cuprate superconductor, which was observed previously in other cuprate families \cite{TALLON200153}.

These findings again show that SuperDiff is not only able to generate new superconductors within known families but is also able to overcome the limitations of previous generative models to generate completely new families of superconductors that are also realistic---although we note that for some reference compounds, SuperDiff was also unable to generate new families of superconductors.

\section{Discussion}
\label{discussion}

With the lack of a systematic approach, the discovery of new high $T_c$ superconductors has long depended on material scientists' serendipity. Recently, machine learning has been applied to this field to help assist scientists, but past works still lacked many key capabilities, for instance, the ability to computationally find new families of superconductors. Moreover, recent generative model approaches applied to this field also lacked methods of controlling the generation process by incorporating information from reference compounds \cite{Kim_2024, wines2023cdvae, zhongdiffsupercon}. 

In this paper, we have introduced a novel method for superconductor discovery using diffusion models with conditioning functionality that has addressed these major issues. Like previous works applying generative models to superconductor discovery, we were able to generate novel, realistic, and highly plausible superconductors that lie outside of existing databases---leveraging this ``inverse design'' approach to significantly outperform manual search and previous classification model approaches. With our unconditional model, we were also able to address the low generated compound uniqueness issues that plagued previous works due to the small training data set for pnictides. Most importantly, however, beyond the unconditional performance improvements the diffusion model brought, our contribution of implementing conditioning with ILVR for superconductor discovery to allow the generation process to be controlled enabled the creation of a tool for computationally generating completely new families of superconductors. We verified the generation of new families of superconductors with our clustering analysis, and we presented several of these promising new families of generated superconductors for several different classes of superconductors in Table \ref{tab:predictions}. Once again, we point out that no previous computational model for superconductor discovery would have been capable of generating these new families of superconductors as they attempt to produce only samples that match the training data.

The application of deep generative models for superconductor discovery continues to be a very promising and exciting approach. Future studies can benefit from possible improvements that can be made to SuperDiff, including implementing a physics-informed diffusion model and creating and utilizing a better, more comprehensive training dataset of superconductors. Nevertheless, SuperDiff in its current form is still very powerful as a tool for superconductor discovery, and researchers can currently benefit from it in a myriad of ways, such as by using its novel generations as inspiration---starting with the new families introduced here, using it to expand on their own new discoveries, or by simply experimenting with many more reference compounds (such as high-pressure superconductors) to continue using it to generate completely new families of hypothetical superconductors or hypothetical superconductors with even higher $T_c$.

\section*{Data availability}

The SuperCon dataset \cite{supercondataset} used in this study to train the SuperDiff model is publicly available at \url{https://doi.org/10.48505/nims.3739}, and a copy of the processed dataset used is available at \url{https://github.com/sdkyuanpanda/SuperDiff/tree/54f0520a67bf8308fbf437b2b66aa36beee52acd/datasets}. The 265,722 valid compounds generated by the four versions of unconditional SuperDiff and the 270 valid compounds generated by conditional SuperDiff conditioned on $\mathrm{YBa_{1.4}Sr_{0.6}Cu_{3}O_{6}Se_{0.51}}$ \cite{grinenko2023extraordinary} are available at \url{https://github.com/sdkyuanpanda/SuperDiff/tree/54f0520a67bf8308fbf437b2b66aa36beee52acd/outputs}. Other data that support the results of this study are available from the corresponding author upon reasonable request.

\section*{Code availability}

An implementation of the proposed model, SuperDiff, is publicly available online at \url{https://github.com/sdkyuanpanda/SuperDiff} and is citable on Zenodo at \url{https://doi.org/10.5281/zenodo.10699906} \cite{yuan_2024_10699906}.

\section*{Additional information}

\textbf{Author contributions statement} Conceptualization and design: S.Y. Data analysis and interpretation: S.Y. and S.V.D. Drafting of the manuscript: S.Y. Critical revision of the manuscript for important intellectual content: S.Y. and S.V.D. Supervision: S.V.D. All authors read and approved the final manuscript.

\vspace{\baselineskip}

\textbf{Competing interests} The authors declare no competing interests.

\bibliography{bibliography}

\end{document}